\documentclass[a4paper, 11pt]{article}
\usepackage{amssymb,amsmath}
\usepackage{graphicx}
\usepackage{latexsym,graphicx}
\usepackage{float}
\usepackage{amsmath}
\usepackage{amsfonts}
\usepackage{amssymb}
\usepackage{epstopdf}
\usepackage{color}
\usepackage {hyperref}
\usepackage{cite}
\usepackage{color}

\usepackage{microtype}

\numberwithin{equation}{section}
\allowdisplaybreaks

\makeatletter \@addtoreset{equation}{section} \makeatother

\usepackage{ytableau}

\ytableausetup{centertableaux, mathmode, smalltableaux}

\definecolor{blue-violet}{rgb}{0.54, 0.17, 0.89}
\definecolor{PineGreen}{cmyk}{0.92, 0, 0.59, 0.25}
\definecolor{YellowOrange}{cmyk}{0, 0.42, 1, 0}

\newcommand{\be}{\begin{equation}}
\newcommand{\ee}{\end{equation}}
\newcommand{\beq} {\begin{equation}}
\newcommand{\eeq} {\end{equation}}
\newcommand{\ba}{\begin{eqnarray}}
\newcommand{\ea}{\end{eqnarray}}

\usepackage{graphicx}

\begin{document}
\title{Myrzakulov $F(T,Q)$ gravity: cosmological implications and constraints}
\author{Dinesh Chandra Maurya$^{1}$\footnote{Email: dcmaurya563@gmail.com}, K. Yesmakhanova$^{2}$\footnote{Email: kryesmakhanova@gmail.com},   
	R. Myrzakulov$^{2}$\footnote{Email: rmyrzakulov@gmai.com} , G. Nugmanova$^{2}$\footnote{Email: nugmanovagn@gmail.com}, \\ 
	$^{1}$\textit{Centre for Cosmology, Astrophysics and Space Science,}\\ \textit{ GLA University, Mathura-281 406, Uttar Pradesh,India}\\
	$^{2}$\textit{Ratbay Myrzakulov Eurasian International Centre for Theoretical} \\ \textit{Physics, Astana 010009, Kazakhstan}\\
}
\date{}
\maketitle
\
\begin{abstract}
\noindent
In this paper, we investigate some exact cosmological models in Myrzakulov $F(T,Q)$  gravity or the  Myrzakulov gravity-III (MG-III) proposed in [arXiv:1205.5266], with observational constraints. The MG-III gravity is some kind of unification of two known gravity theories, namely, the $F(T)$ gravity and the $F(Q)$ gravity. The field equations of the MG-III theory are obtained by regarding the metric tensor and the general affine connection as independent variables. We then focus on the particular case in which the $F(T,Q)$ function characterizing the aforementioned metric-affine models is linear that is $F(T,Q)=\lambda T+\mu Q$. We investigate this linear case and consider a Friedmann-Lema\^{i}tre-Robertson-Walker background to study cosmological aspects and applications. We have obtained three exact solutions of the modified field equations in different cases $T$ and $Q$, in the form of Hubble function $H(t)$ and scale factor $a(t)$ and placed observational constraints on it through the Hubble $H(z)$ datasets on it using the MCMC analysis. We have investigated the deceleration parameter $q(z)$, effective EoS parameters and a comparative study of all three models with $\Lambda$CDM model has been carried out.\\

\end{abstract}
	\vspace{1cm}

\tableofcontents

\section{Introduction}

The discovery of the Universe's unexpectedly accelerating expansion \cite{ref1,ref2,ref3,ref4,ref5}, which began at a redshift $z$ of approximately $z\approx0.6$, has raised doubts about the fundamental principles of general relativity, the most successful theory of gravity to date. To elucidate the late-time acceleration, the most straightforward approach is to revert back to the original cosmological constant $\Lambda$ that Einstein used in order to develop the initial general relativistic cosmic model \cite{ref6}. The $\Lambda$CDM paradigm, which is the standard model of cosmology, was formulated using this assumption and is substantially corroborated by the exceptional agreement between the cosmological constant and the observed data. This paradigm also requires the presence of another fascinating (but currently unknown) element of the Universe, referred to as dark matter \cite{ref7}. The $\Lambda$CDM model effectively accounts for observed phenomena \cite{ref8,ref9,ref10,ref11}; yet, it encounters a notable challenge: there is no underlying basic physical theory that can provide an explanation for it. The main problem arises from the difficulties in explaining the characteristics and origin of the cosmological constant \cite{ref12,ref13,ref14}.\\

Recent observational data has raised doubts about the status of General Relativity (GR), despite its unquestionable elegance and effectiveness as a theory in physics \cite{ref15}. The rapid expansion of our Universe during both its early and late stages, which cannot be explained within the framework of General Relativity, is arguably the most noteworthy observation. Due to the gap between theory and observations, several theories other than General Relativity (GR) have been formulated. These theories, known as Modified Gravity, aim to address this issue \cite{ref16}. The hunt of a feasible alternative has shown to be advantageous and productive in enhancing our understanding of gravity. Examples of the abundance of modified gravity theories include metric $f(R)$ theories, Metric-Affine (Palatini) $f(R)$ gravity \cite{ref17, ref18, ref19}, teleparallel $f(T)$ gravity theories \cite{ref20, ref21}, symmetric teleparallel $f(Q)$ gravity theories \cite{ref22, ref23}, and Scalar-Tensor theories \cite{ref24, ref25}.\\

The linear representations of $f(T)$ result in a teleparallel gravity theory that is comparable to general relativity (TEGR) \cite{ref26}. However, there are differences in the way the two theories of gravity, $f(T)$ and $f(R)$, are understood in terms of their physical interpretations. In $f(T)$ gravity, the torsion scalar $T$ is defined by the first-order derivatives of the vierbeins. However, in $f(R)$ gravity, the Ricci scalar $R$ includes the second-order derivatives of the metric tensor. Unlike other modified theories of gravity, the cosmological models in $f(T)$ gravity allow for easily finding the exact solutions. The $f(T)$ gravity theory is a simplified and modified version of gravity. However, there is a lack of accurate answers proposed in the existing literature. Power-law solutions have been found in the literature for cosmological models in both isotropic and anisotropic spacetime, as evidenced by \cite{ref27,ref28,ref29}. Cosmologists have examined precise solutions of cosmological models in \cite{ref30,ref31}. These solutions pertain to static spherically symmetric spacetime and Bianchi type-I spacetime. The analysis of cosmic conditions in $f(T)$ gravity is more straightforward than in previous modified theories of gravity. Therefore, numerous cosmological scenarios, including as the huge bounce \cite{ref32, ref33, ref34, ref35}, inflationary model \cite{ref36}, and late time cosmic acceleration \cite{ref37, ref38, ref39}, are investigated utilizing the $f(T)$ gravity theory. Recent advancements in the realm of $f(T)$ gravity encompass the discovery of spherical and cylindrical solutions \cite{ref40}, the invention of conformally symmetric traversable wormholes \cite{ref41}, and the exploration of noether charge and black hole entropy \cite{ref42}. We have recently examined and reconstructed several $\Lambda$CDM cosmological models within the framework of $f(T)$ gravity, as described in \cite{ref43,ref44,ref45,ref46}.\\

On the other hand, the investigation of the gravitational interaction mediated by non-metricity, in the absence of curvature and torsion, is a recent and fascinating alternative that has been explored in recent studies \cite{ref47,ref48,ref49,ref50,ref51,ref52,ref53}. This approach is essential for explaining the fundamental nature of gravity, as it allows us to interpret gravity as a gauge theory without explicitly assuming the validity of the Equivalence Principle. Examining the $f(Q)$ theories, where $Q$ represents the non-metricity scalar, in this particular situation can offer novel insights into the cosmic acceleration that arises from the inherent consequences of an alternative geometry to the Riemannian geometry. The analysis of the connecting matter in $f(Q)$ gravity assumes a power-law function, as stated in \cite{ref54}. Recently, in \cite{ref55}, a model-independent reconstruction approach was employed to investigate cosmological properties. The paper \cite{ref56} presents a formulation of general relativity and its scalar-tensor extension using nonmetricity. Additionally, the paper \cite{ref57} explores general relativity with spin and torsion. The article by \cite{ref58} presents a Covariant formulation of the $f(Q)$ theory, while \cite{ref59} suggests the $f(Q, T)$ theory as an expansion of $f(Q)$ gravity. A recent article by \cite{ref60} provides a comprehensive assessment of different cosmological theories in the context of $f(Q)$ gravity. The paper \cite{ref61} examines the wormhole geometry in $f(Q)$ gravity while considering energy conditions. On the other hand, \cite{ref62} explores the Tsallis holographic dark energy in viscous $f(Q)$ gravity by using the tachyon field. The paper \cite{ref63} presents an analysis of a dynamical system in $f(Q)$ gravity with perturbation. Additionally, the papers \cite{ref64,ref65,ref66} study string-fluid cosmological models in $f(Q)$ gravity. The authors of references \cite{ref67,ref68,ref69,ref70,ref71} have recently studied transit phase accelerating cosmological models with observational constraints in the context of the extension of $f(Q)$ gravity.\\

Undoubtedly, the selection of modifications is primarily determined by individual taste. In our viewpoint, captivating and deeply driven alternatives are those that offer a broader correlation than the conventional Levi-Civita connection, thereby expanding the fundamental geometry of spacetime. Under typical conditions, the space will exhibit non-Riemannian \cite{ref72} properties, such as torsion and non-metricity, when the connection is not limited beforehand and is seen as an additional basic field alongside the metric. Once the affine relationship has been determined, these last geometric quantities can be computed. Metric-Affine theories of gravity are theories that have been formulated on a non-Riemannian manifold \cite{ref73,ref74}.\\

The Metric-Affine approach \cite{ref19,ref75,ref76,ref77,ref78,ref79,ref80,ref81,ref82,ref83,ref84,ref85,ref86,ref87,ref88,ref89} has gained considerable prominence in recent years, especially for its applications in cosmology \cite{ref90,ref91,ref92,ref93,ref94,ref95,ref96,ref97,ref98,ref99,ref100,ref101}. The interest in this matter may arise from the clear geometric interpretation of the supplementary influences that function within this framework, as opposed to general relativity (GR). Simply put, the changes can be attributed solely to spacetime torsion and non-metricity. Furthermore, the presence of matter possessing inherent structure stimulates these geometric notions, as referenced in \cite{ref92,ref102,ref103,ref104,ref105}. The MAG scheme benefits from the connection between generalized geometry and inner structure, which adds another advantageous component. For this investigation, we shall consider this framework.\\

These theories serve as impetus for developing a theory based on affinely connected metrics, namely the Riemann-Cartan subclass \cite{ref106}, utilizing a specific but non-special connection. This would offer the additional degrees of freedom commonly required in any gravitational modification by simultaneously generating both non-zero curvature and non-zero torsion \cite{ref107}. Therefore, the explanation of the evolution of both the early and late universe may be adequately provided by Myrzakulov gravity, as stated in \cite{ref108,ref109,ref110,ref111,ref112}. The reference \cite{ref108} presents a recent investigation on the cosmology that arises from employing a certain framework. The study involves calculating the changes in observable quantities, such as the density parameters and the effective dark energy equation-of-state parameter, over time. The researchers have investigated the cosmic behavior by focusing on the impact of connections. They have employed the mini-super-space technique to represent the theory as a deformation of both general relativity and its teleparallel counterpart. The study conducted by \cite{ref113} has examined the observational constraints of Myrzakulov $F(R,T)$-gravity. The paper \cite{ref114,ref115,ref116,ref117,ref118,ref119,ref120} explores many Metric-Affine Myrzakulov Gravity Theories and their practical uses.\\

We have recently examined cosmological models in Myrzakulov $F(R,T)$ gravity, taking into account observational constraints \cite{ref121}. Additionally, specific cosmological models inside this metric-affine $F(R,T)$ gravity have been investigated in \cite{ref122}. In \cite{ref123}, we have investigated certain precise cosmological models and their features in the Metric-Affine Myrzakulov Gravity-II, often known as the $F(R,Q)$ gravity theory, inspired by the aforementioned talks. In this paper, we will examine the function $F(T,Q)=\lambda T+\mu Q$, where $T$ represents the torsion scalar, $Q$ represents the non-metricity scalar with regard to a non-special connection, and $\mu$ is an arbitrary constant.\\

The present paper is organized in the following sections: Sect.-2 contains some geometrical concepts of metric-affine spacetime, and a brief introduction of the Myrzakulov $F(T,Q)$ gravity is given in sect.-3. Sect.-4 deals with cosmological field equations of $F(T,Q)$ gravity in a flat FLRW spacetime and we have obtained deducted gravity field equations from $F(T,Q)$ gravity theory. In Sect.-5, we obtained some exact solutions of the derived field equations in different choices of $v$ and $w$. We have made observational constraints on the models obtained using Hubble datasets $H(z)$ by applying MCMC analysis in Sect.-6. Result discussions are explored in Sect.-7, and finally conclusions are given in last Sect.-8. 

\section{Geometrical preliminaries}

We take into consideration the metric-affine spacetime, which is a generic spacetime featuring nonmetricity, torsion, and curvature. The connection in this spacetime is described as
\begin{equation}\label{2.1}
	\Gamma^{\rho}_{\,\,\,\,\mu\nu}=\breve{\Gamma}_{\,\,\,\, \mu \nu}^{\rho}+K^{\rho}_{\,\,\,\,\mu\nu}+L^{\rho}_{\,\,\,\,\mu\nu}\,,
\end{equation}
where the Levi--Civita connection is denoted by $\breve{\Gamma}_{\,\,\,\, \mu \nu}^{\rho}$, the contorsion tensor is denoted by $K^{\rho}_{\,\,\,\,\mu\nu}$, and the disformation tensor is denoted by $L^{\rho}_{\,\,\,\,\mu\nu}$. The following are the forms of these three tensors.
\begin{eqnarray}
	\breve{\Gamma}^l_{\, \, \, jk} &=& \tfrac{1}{2} g^{lr} \left( \partial _k g_{rj} + \partial _j g_{rk} - \partial _r g_{jk} \right)\,, \label{2.2}\\
	K^{\rho}_{\,\,\,\,\mu\nu}&=&\frac{1}{2}g^{\rho\lambda}\bigl(T_{\mu\lambda\nu}+T_{\nu\lambda\mu}+T_{\lambda\mu\nu}\bigr)
	=-K^{\rho}_{\,\,\,\,\nu\mu}\, , \label{2.3}	\\	
	L^{\rho}_{\,\,\,\,\mu\nu}&=&\frac{1}{2}g^{\rho\lambda}\bigl(-Q_{\mu \nu \lambda}-Q_{\nu \mu \lambda} + Q_{\lambda\mu\nu}\bigr)=
	L^{\rho}_{\,\,\,\,\nu\mu}. \label{2.4}
\end{eqnarray}
Here \begin{equation}
	T_{\,\,\,\, \mu \nu}^{\alpha}=2 \Gamma_{\,\,\,\, [\mu \nu]}^{\alpha}, \quad Q_{\rho \mu \nu} = \nabla_{\rho} g_{\mu \nu}\,, \label{2.5}
\end{equation}
The torsion tensor and the nonmetricity tensor are distinct mathematical quantities. Within this overarching spacetime framework, which encompasses curvature, torsion, and nonmetricity, we shall now define the following three tensors:
\begin{eqnarray}
	R_{jk}&=&\partial_i\Gamma_{jk}^i-\partial_j\Gamma_{ik}^i+\Gamma_{ip}^i\Gamma_{jk}^p-\Gamma_{jp}^i\Gamma_{ik}^p\,, \label{2.6} \\
	S^{p\mu\nu}&=&K^{\mu\nu p}-g^{p\nu}T^{\sigma\mu}_{\,\,\sigma}+g^{p\mu}T^{\sigma\nu}_{\,\,\sigma}\,, \label{2.7} \\
	K^{\nu}_{p\mu}&=&\frac{1}{2}(T_{p\,\,\,\mu}^{\,\,\nu}+T_{\mu\, \, \, p}^{\,\, \nu}-T^{\nu}_{p\mu}). \label{2.8}
\end{eqnarray}
The Ricci tensor, the potential, and the contorsion tensor are denoted by these terms, respectively. Next, we will present three geometric scalars as
\begin{eqnarray}
	R&=&g^{\mu\nu}R_{\mu\nu}\,,\label{2.9}\\
	T&=&{S_\rho}^{\mu\nu}\,{T^\rho}_{\mu\nu}\,,\label{2.10}\\
	Q&=& -g^{\mu\nu}(L^{\alpha}_{\beta\mu}L^{\beta}_{\nu\alpha}-L^{\alpha}_{\beta\alpha}L^{\beta}_{\mu\nu})\,, \label{2.11}
\end{eqnarray}  
The variables $R$, $T$, and $Q$ represent the curvature scalar, torsion scalar, and nonmetricity scalar, respectively. Based on the framework we described in reference [124], we make the assumption that these three scalars can be expressed in the following forms. 
\begin{eqnarray}
	R&=&u+R_{s}\,, \label{2.12}\\
	T&=&v+T_{s}\,, \label{2.13}\\
	Q&=&w+Q_{s}\,, \label{2.14}
\end{eqnarray}
where $u=u(\Gamma^{\rho}_{\,\,\,\,\mu\nu}, x_i, g_{ij}, \dot{g_{ij}},\ddot{g_{ij}}, ... , f_j), \quad v=v(\Gamma^{\rho}_{\,\,\,\,\mu\nu}, x_i, g_{ij}, \dot{g_{ij}},\ddot{g_{ij}}, ... , g_j)$  and $w=w(\Gamma^{\rho}_{\,\,\,\,\mu\nu}, x_i, \\  g_{ij}, \dot{g_{ij}},\ddot{g_{ij}}, ... , h_j)$ are some real functions. Here: i) $R_{s}=R^{(LC)}$ is the curvature scalar corresponding to the Levi-Civita connection with the vanishing torsion and nonmetricity ($T=Q=0$); ii) $T_{s}=T^{(WC)}$ is the torsion scalar for the purely Weitzenb\"{o}ck connection with the vanishing curvature and nonmetricity ($R=Q=0$); iii) $Q_{s}=Q^{(NM)}$ is the nonmetricity scalar with the vanishing torsion and curvature ($R=T=0$).

\section{Myrzakulov $F(T,Q)$ gravity or Myrzakulov gravity-III}

In this paper, we consider the Myrzakulov $F(T,Q)$ gravity or Myrzakulov gravity-III (MG-III)  theory  \cite{ref124}
\begin{equation}
\mathcal{S} = \frac{1}{2 \kappa} \int \sqrt{-g} d^4 x \left[ F(T,Q) + 2 \kappa \mathcal{L}_{\text{m}} \right] \,. \label{3.1}
\end{equation}
In this model, the function $F=F(T,Q)$ is a generic function of the torsion scalar $T$ and the non-metricity scalar $Q$. The action \eqref{3.1} is an extension of both the $F(T)$ and $F(Q)$ theories. This means the MG-III is the unification of  $F(T)$ and $F(Q)$ gravity theories. 
Varying \eqref{3.1} with respect to the metric field we get
\begin{equation}
\begin{split}
& - \frac{1}{2} g_{\mu \nu} F + F_T \left( 2 S_{\nu \alpha \beta} {S_\mu}^{\alpha \beta} - S_{\alpha \beta \mu} {S^{\alpha \beta}}_\nu + 2 S_{\nu \alpha \beta} {S_\mu}^{\beta \alpha} - 4 S_\mu S_\nu \right) + F_Q L_{(\mu \nu)} \\
& + \hat{\nabla}_\lambda \left(F_Q {J^\lambda}_{(\mu \nu)} \right) + g_{\mu \nu} \hat{\nabla}_\lambda \left(F_Q \zeta^\lambda \right) = \kappa T_{\mu \nu} \,, \label{3.2}
\end{split}
\end{equation}
 where $F_{T}=\frac{\partial F}{\partial T}$, $F_{Q}=\frac{\partial F}{\partial Q}$ and $T_{\mu \nu }=-\frac{2}{\sqrt{-g}}\frac{\delta \left( \sqrt{-g}\mathcal{L}_{m}\right) }{\delta g^{\mu \nu }}$,  
\begin{equation}
	\hat{\nabla}_\lambda := \frac{1}{\sqrt{-g}} \left( 2 S_\lambda - \nabla_\lambda \right)\,, \label{3.3}
\end{equation}
and 
\begin{equation}\label{3.4}
	\begin{split}
		L_{\mu \nu} & := \frac{1}{4} \left[ \left(Q_{\mu \alpha \beta} - 2 Q_{\alpha \beta \mu} \right) {Q_\nu}^{\alpha \beta} + \left( Q_\mu + 2 \tilde{Q}_\mu \right) Q_\nu + \left( 2 Q_{\mu \nu \alpha} - Q_{\alpha \mu \nu} \right) Q^\alpha \right] \\
		& \phantom{:= \,} - {\Xi^{\alpha \beta}}_\nu Q_{\alpha \beta \mu} - \Xi_{\alpha \mu \beta} {Q^{\alpha \beta}}_\nu \,, \\
		{J^\lambda}_{\mu \nu} & := \sqrt{-g} \left(\frac{1}{4} {Q^\lambda}_{\mu \nu} - \frac{1}{2} {Q_{\mu \nu}}^\lambda + {\Xi^\lambda}_{\mu \nu} \right) \,, \\
		\zeta^\lambda & := \sqrt{-g} \left( - \frac{1}{4} Q^\lambda + \frac{1}{2} \tilde{Q}^\lambda \right) \,, 
	\end{split}
\end{equation}
where $Q_{\lambda \mu \nu}$ is the non-metricity tensor, $Q_\lambda$ and $\tilde{Q}_\lambda$ are its trace parts, and $\Xi_{\lambda \mu \nu}$ is the so-called (non-metricity) ``superpotential".  We
assume that the matter is a perfect fluid whose energy-momentum tensor $T_{\mu \nu }$ is given by 
\begin{equation}\label{3.5}
	T_{\mu \nu }=(\rho +p)u_{\mu }u_{\nu }+pg_{\mu \nu }\,,
\end{equation}%
where $u_{\mu }$ is the four-velocity satisfying the normalization condition
$u_{\mu }u^{\mu }=-1$, $\rho $ and $p$ are the energy density and pressure
of a perfect fluid respectively.\\
On the other hand, from the variation of \eqref{3.1} with respect to the general affine connection ${\Gamma^\lambda}_{\mu \nu}$ we obtain
\begin{equation}\label{3.6}
2 F_T \left({S^{\mu \nu}}_\lambda - 2 {S_\lambda}^{[\mu \nu]} - 4 S^{[\mu} \delta^{\nu]}_\lambda \right) + F_Q \left[ 2 {Q^{[\nu \mu]}}_\lambda - {Q_\lambda}^{\mu \nu} + \left( q^\nu - Q^\nu \right) \delta^\mu_\lambda + Q_\lambda g^{\mu \nu} + \frac{1}{2} Q^\mu \delta^\nu_\lambda \right] = 0 \,.
\end{equation}

\section{FLRW cosmological field equations of $F(T,Q)$ gravity}

First, let us rewrite the action \eqref{3.1} as 
\begin{eqnarray}\label{4.1}
	S=\frac{1}{2\kappa^{2}}\int \sqrt{-g}d^{4}x[F(T,Q)-\lambda_{1}(T-T_{s}-v)-\lambda_{3}(Q-Q_{s}-w)+2\kappa^{2}L_{m}].
\end{eqnarray}
The variations of the action with respect to $T, Q$   give $\lambda_{1} = F_{T}, \lambda_{3}=F_{Q}$, respectively. Thus the action of the MG-III  takes the form
\begin{eqnarray}\label{4.2}
	S=\frac{1}{2\kappa^{2}}\int \sqrt{-g}d^{4}x[F-F_{T}(T-T_{s}-v)-F_{Q}(Q-Q_{s}-w)+2\kappa^{2}L_{m}],
\end{eqnarray}
where we assume that $v=v(g_{ij}, \dot{g}_{ij}, \ddot{g}_{ij}, ...),  \quad w=w(g_{ij}, \dot{g}_{ij}, \ddot{g}_{ij}, ...)$.
We now consider the FLRW spacetime case with the metric 
\begin{equation}\label{4.3}
	ds^2=-N^{2} (t)dt^2+a^2(t)(dx^2+dy^2+dz^2),
\end{equation}
where $a=a(t)$ is the scale factor, $N(t)$ is the lapse function and taking $N(t)=1$. Then integrating by parts gives the following action with the point-like FLRW Lagrangian 
\begin{eqnarray}\label{4.4}
	S=\frac{1}{2\kappa^{2}}\int {\cal L}dt,
\end{eqnarray}
where the point-like Lagrangian has the form
\begin{eqnarray}\label{4.5}
	{\cal L}= a^{3} [F-F_{T}(T-T_{s}-v)-F_{Q}(Q-Q_{s}-w)+2\kappa^{2}L_{m}].
\end{eqnarray}
In FLRW spacetime, we have
\begin{eqnarray}
	T_{s}&=&-6H^{2}\,, \label{4.6}\\
	Q_{s}&=&6H^{2}. \label{4.7}
\end{eqnarray}
Finally we get the following FLRW Lagrangian
\begin{eqnarray}\label{4.8}
	{\cal L}(a,T,Q,\dot{a},\dot{T},\dot{Q})= a^{3}(F-TF_{T}-QF_{Q})-6a\dot{a}^{2}(F_{T}-F_{Q})+a^{3}(vF_{T}+wF_{Q})+2\kappa^{2}a^{3}L_{m},
\end{eqnarray}
Now, taking the Hamiltonian ${\cal H}$ of Lagrangian ${\cal L}$ as 
\begin{eqnarray}\label{4.9}
	{\cal H}={\cal E}=\dot{a}\frac{\partial {\cal L}}{\partial \dot{a}}+\dot{T}\frac{\partial {\cal L}}{\partial \dot{T}}+\dot{Q}\frac{\partial {\cal L}}{\partial \dot{Q}}-{\cal L}=0
\end{eqnarray}
and the Euler-Lagrange equations corresponding the Lagrangian ${\cal L}$, we obtain the following field equations
\begin{equation}\label{4.10}
	-\frac{1}{2}(F-TF_{T}-QF_{Q})-3H^{2}(F_{T}-F_{Q})-\frac{1}{2}[(v-\dot{a}v_{\dot{a}})F_{T}+(w-\dot{a}w_{\dot{a}})F_{Q}]=\kappa^{2}\rho
\end{equation}
\begin{multline}\label{4.11}
	-\frac{1}{2}(F-TF_{T}-QF_{Q})-(2\dot{H}+3H^{2})(F_{T}-F_{Q})-\frac{1}{2}(v+\frac{1}{3}av_{a}-\dot{a}v_{\dot{a}}-\frac{1}{3}a\dot{v}_{\dot{a}})F_{T}\\-\frac{1}{2}(w+\frac{1}{3}aw_{a}-\dot{a}w_{\dot{a}}-\frac{1}{3}a\dot{w}_{\dot{a}})F_{Q}-2H(\dot{F}_{T}-\dot{F}_{Q})+\frac{1}{6}a(u_{\dot{a}}\dot{F}_{T}+w_{\dot{a}}\dot{F}_{Q})=-\kappa^{2}p,
\end{multline}
where
\begin{equation}\label{4.12}
	\rho=L_{m}-\dot{a}\frac{\partial L_{m}}{\partial \dot{a}},~~~~p=\frac{1}{3a^{2}}\left[\frac{d}{dt}\left(a^{3}\frac{\partial L_{m}}{\partial \dot{a}}\right)-\frac{\partial}{\partial a}(a^{3}L_{m})\right].
\end{equation}

\subsection{Case 1: $F=-T$ gravity}

We can derive various gravity theories from this generalized metric-affine $F(T,Q)$ gravity theory, first, we consider the case $F=-T$, then Eqs.~\eqref{4.10} \& \eqref{4.11} becomes
\begin{eqnarray}
	3H^{2}+\frac{1}{2}(v-\dot{a}v_{\dot{a}})&=&\kappa^{2}\rho\,,  \label{4.13} \\
	2\dot{H}+3H^{2}+\frac{1}{2}(v+\frac{1}{3}av_{a}-\dot{a}v_{\dot{a}}-\frac{1}{3}a\dot{v}_{\dot{a}})&=&-\kappa^{2}p.  \label{4.14}
\end{eqnarray}
Obviously, if we take $v=0$, we obtain the standard form of Friedmann equations
\begin{eqnarray}
	3H^{2}&=&\kappa^{2}\rho\,,  \label{4.15} \\
	2\dot{H}+3H^{2}&=&-\kappa^{2}p.  \label{4.16}
\end{eqnarray}

\subsection{Case 2: $F(R)$ gravity}

Now we consider the case  $F=F(T)$, then Eqs.~\eqref{4.10} \& \eqref{4.11} becomes
\begin{equation}
		-\frac{1}{2}(F-TF_{T})-3H^{2}F_{T}-\frac{1}{2}(v-\dot{a}v_{\dot{a}})F_{T}=\kappa^{2}\rho\,,  \label{4.17}
\end{equation}
\begin{equation}
	-\frac{1}{2}(F-TF_{T})-(2\dot{H}+3H^{2})F_{T}-\frac{1}{2}(v+\frac{1}{3}av_{a}-\dot{a}v_{\dot{a}}-\frac{1}{3}a\dot{v}_{\dot{a}})F_{T}-2H\dot{F}_{T}+\frac{1}{6}av_{\dot{a}}\dot{F}_{T}=-\kappa^{2}p.  \label{4.18}
\end{equation}
If we put $v=0$, then we obtain the $F(T)$ gravity field equations in Riemann-manifold as
\begin{eqnarray}
	-\frac{1}{2}(F-TF_{T})-3H^{2}F_{T}&=&\kappa^{2}\rho,  \label{4.19} \\
	-\frac{1}{2}(F-TF_{T})-(2\dot{H}+3H^{2})F_{T}-2H\dot{F}_{T}&=&-\kappa^{2}p.  \label{4.20}
\end{eqnarray}

\subsection{Case 3: $F=Q$ gravity}

If we consider the case $F=Q$, then Eqs.~\eqref{4.10} \& \eqref{4.11} becomes
\begin{eqnarray}
	3H^{2}-\frac{1}{2}(w-\dot{a}w_{\dot{a}})&=&\kappa^{2}\rho,  \label{4.21} \\
	2\dot{H}+3H^{2}-\frac{1}{2}(w+\frac{1}{3}aw_{a}-\dot{a}w_{\dot{a}}-\frac{1}{3}a\dot{w}_{\dot{a}})&=&-\kappa^{2}p\,,  \label{4.22}
\end{eqnarray}
If we assume $w=0$, then we get the following system of equations as usual
\begin{eqnarray}
	3H^{2}&=&\kappa^{2}\rho,  \label{4.23} \\
	2\dot{H}+3H^{2}&=&-\kappa^{2}p.  \label{4.24}
\end{eqnarray}

\subsection{Case 4: $F(Q)$ gravity}

If we take the case $F=F(Q)$, then we get the metric-affine $F(Q)$ gravity field equations as
\begin{equation}
		-\frac{1}{2}(F-QF_{Q})+3H^{2}F_{Q}-\frac{1}{2}(w-\dot{a}w_{\dot{a}})F_{Q}=\kappa^{2}\rho\,,  \label{4.25} 
\end{equation}
\begin{equation}
	-\frac{1}{2}(F-QF_{Q})+(2\dot{H}+3H^{2})F_{Q}-\frac{1}{2}(w+\frac{1}{3}aw_{a}-\dot{a}w_{\dot{a}}-\frac{1}{3}a\dot{w}_{\dot{a}})F_{Q}+2H\dot{F}_{Q}+\frac{1}{6}aw_{\dot{a}}\dot{F}_{Q}=-\kappa^{2}p\,,  \label{4.26}
\end{equation}
If we assume $w=0$, then we get the following system of equations as usual
\begin{eqnarray}
	-\frac{1}{2}(F-QF_{Q})+3H^{2}F_{Q}&=&\kappa^{2}\rho,  \label{4.27} \\
	-\frac{1}{2}(F-QF_{Q})+(2\dot{H}+3H^{2})F_{Q}+2H\dot{F}_{Q}&=&-\kappa^{2}p.  \label{4.28}
\end{eqnarray}
Thus, we have seen that the MG-III $F(T,Q)$ gravity theory is the generalization of both $F(T)$ and $F(Q)$ gravity theories.

\section{Cosmological solutions for $F(T,Q)=\lambda T+\mu Q$ gravity. }

We possess a pair of field equations \eqref{4.10} and \eqref{4.11} that are not dependent on each other, and these equations involve six unknowns: $F, \rho, p, a, v, w$. In order to obtain precise solutions for these two field equations, it is necessary to impose a minimum of four restrictions on these unknown variables. Thus, we begin by examining the specific scenario of the arbitrary function $F(T,Q)= \lambda T+\mu Q$, where $\lambda$ and $\mu$ are model parameters. Subsequently, the field equations \eqref{4.10} and \eqref{4.11} are modified accordingly.
\begin{equation}
	3(\mu-\lambda)H^{2}-\frac{1}{2}[\lambda(v-\dot{a}v_{\dot{a}})+\mu(w-\dot{a}w_{\dot{a}})]=\kappa^{2}\rho,  \label{5.1} 
\end{equation}
\begin{equation}
	(\mu-\lambda)(2\dot{H}+3H^{2})-\frac{1}{2}[\lambda(v+\frac{1}{3}av_{a}-\dot{a}v_{\dot{a}}-\frac{1}{3}a\dot{v}_{\dot{a}})+\mu(w+\frac{1}{3}aw_{a}-\dot{a}w_{\dot{a}}-\frac{1}{3}a\dot{w}_{\dot{a}})]=-\kappa^{2}p,  \label{5.2}
\end{equation}
Simultaneously, the continuity equation is expressed in the following manner for the initial density and pressure.
\begin{eqnarray}
	\dot{\rho}+3H(\rho+ p)+\frac{1}{2\kappa^{2}}(\dot{y}-\dot{a}y_{a}-\ddot{a}y_{\dot{a}})=0,  \label{5.3}
\end{eqnarray}
where
\begin{eqnarray}
	y=\lambda v+\mu w.  \label{5.4}
\end{eqnarray}
The modified $F(T,Q)$ gravity theory is contingent upon the selection of the parameters $v$ and $w$, which can be determined based on their respective definitions. Hence, we examine the aforementioned model to determine the optimal values of $v=k_{1}H^{m}$ and $w=k_{2}H^{n}$, resulting in the subsequent precise cosmological models outlined below.

By plugging in the given values of $v$ and $w$ into equations \eqref{5.1}, \eqref{5.2} \& \eqref{5.3}, we can calculate the result as follows
\begin{equation}
	3(\mu-\lambda)H^{2}-\frac{1}{2}[\lambda k_{1}(1-m)H^{m}+\mu k_{2}(1-n)H^{n}]=\kappa^{2}\rho\,, \label{5.5}
\end{equation}
\begin{multline}
	(\mu-\lambda)(2\dot{H}+3H^{2})-\frac{\lambda k_{1}}{2}\left[\frac{3-5m}{3}H^{m}-\frac{m(m-1)}{3}H^{m-2}\dot{H}\right]\\-\frac{\mu k_{2}}{2}\left[\frac{3-5n}{3}H^{n}-\frac{n(n-1)}{3}H^{n-2}\dot{H}\right]=-\kappa^{2}p\,, \label{5.6}
\end{multline}
\begin{equation}
	\dot{\rho}+3H(\rho+ p)=0\,, \label{5.7}
\end{equation}
Now, we want to find exact solutions of these two field equations \eqref{5.5} \& \eqref{5.6} for the different values of $m$ and $n$, and investigate the cosmological implications and constraints on each solutions.

\subsection{Model-I for $m=n=0$}

For $m=n=0$, the above field equations \eqref{5.5} \& \eqref{5.6} reduces to
\begin{equation}
	3(\mu-\lambda)H^{2}-\frac{\lambda k_{1}+\mu k_{2}}{2}=\kappa^{2}\rho\,, \label{5.8}
\end{equation}
\begin{equation}
	(\mu-\lambda)(2\dot{H}+3H^{2})-\frac{\lambda k_{1}+\mu k_{2}}{2}=-\kappa^{2}p\,, \label{5.9}
\end{equation}
Now, we define the equation of state (EoS) for the perfect fluid considered as $p=\omega\rho$ with the assumption $\omega$ the EoS parameter as constant. Using this constraints in Eqs.~\eqref{5.8} \& \eqref{5.9}, we obtain the following equation
\begin{equation}
	\dot{H}+\frac{3(1+\omega)}{2}H^{2}-\frac{(1+\omega)(\lambda k_{1}+\mu k_{2})}{4(\mu-\lambda)}=0\,, \label{5.10}
\end{equation}
After integration above equation \eqref{5.10} with respect to cosmic time $t$, we get the Hubble function $H(t)$ as
\begin{equation}
	H(t)=\sqrt{\frac{6(\mu-\lambda)}{\lambda k_{1}+\mu k_{2}}} \coth\left(\frac{1+\omega}{2}\sqrt{\frac{3(\lambda k_{1}+\mu k_{2})}{2(\mu-\lambda)}}t+c_{0}\sqrt{\frac{6(\mu-\lambda)}{\lambda k_{1}+\mu k_{2}}}\right)\,, \label{5.11} 
\end{equation}
and the corresponding scale factor is obtained as
\begin{equation}
	a(t)=c_{1}\left[\sinh\left(\frac{1+\omega}{2}\sqrt{\frac{3(\lambda k_{1}+\mu k_{2})}{2(\mu-\lambda)}}t+c_{0}\sqrt{\frac{6(\mu-\lambda)}{\lambda k_{1}+\mu k_{2}}}\right)\right]^{\frac{2}{3(1+\omega)}}\,, \label{5.12}
\end{equation}
where $c_{0}, c_{1}$ are arbitrary constants of integration.\\
Now, we can rewrite the field equations Eqs.~\eqref{5.8} \& \eqref{5.9} in the standard Friedmann field equations as
\begin{equation}
	3H^{2}=\kappa^{2}\rho+\kappa^{2}\rho_{(geom)}, \label{5.13}
\end{equation}
\begin{equation}
	2\dot{H}+3H^{2}=-\kappa^{2}p-\kappa^{2}p_{(geom)}, \label{5.14}
\end{equation}
where the effective dark energy density and pressure $\rho_{(geom)}$ and $p_{(geom)}$ coming from geometrical modifications are as given below:
\begin{equation}
	\rho_{(geom)}=\frac{1}{\kappa^{2}}\left[\frac{\lambda k_{1}+\mu k_{2}}{2}+3(1-\mu+\lambda)H^{2}\right]\,, \label{5.15}
\end{equation}
\begin{equation}
	p_{(geom)}=-\frac{1}{\kappa^{2}}\left[\frac{\lambda k_{1}+\mu k_{2}}{2}+(1-\mu+\lambda)(2\dot{H}+3H^{2})\right]\,, \label{5.16}
\end{equation}
and hence, the effective dark EoS parameter $\omega_{(geom)}$ is defined as $p_{(geom)}=\omega_{(geom)}\rho_{(geom)}$ and obtained as
\begin{equation}
	\omega_{(geom)}=-1-\frac{2(1-\mu+\lambda)\dot{H}}{\frac{\lambda k_{1}+\mu k_{2}}{2}+3(1-\mu+\lambda)H^{2}}\,, \label{5.17}
\end{equation}
Now, solving Eq.~\eqref{5.7}, we get the matter energy density $\rho$ as
\begin{equation}
	\rho=\rho_{0}\left(\frac{a_{0}}{a}\right)^{3(1+\omega)}\,, \label{5.18}
\end{equation}
where $\rho_{0}$ is an integrating constant.\\
Using the relationship of scale factor and redshift $z$, $\frac{a_{0}}{a}=1+z$ with $a_{0}=1$ for standard convention, we have
\begin{equation}
	\rho(z)=\rho_{0}(1+z)^{3(1+\omega)}\,, \label{5.19}
\end{equation}
\begin{equation}
	H(z)=\sqrt{\frac{6(\mu-\lambda)}{\lambda k_{1}+\mu k_{2}}}\sqrt{1+[c_{1}(1+z)]^{3(1+\omega)}}\,, \label{5.20}
\end{equation}
Also, re-writing Eq.~\eqref{5.20}, as
\begin{equation}
	H(z)=\frac{H_{0}}{\sqrt{1+c_{1}^{3(1+\omega)}}}\sqrt{1+[c_{1}(1+z)]^{3(1+\omega)}}\,, \label{5.21}
\end{equation}
where $H_{0}=\sqrt{\frac{6(\mu-\lambda)}{\lambda k_{1}+\mu k_{2}}}\sqrt{1+c_{1}^{3(1+\omega)}}$, the value $H(z)$ at $z=0$.\\
Now, we have obtained the deceleration parameter $q(z)=-1+(1+z)\frac{H'}{H}$ with $H'=\frac{dH}{dz}$, as
\begin{equation}
	q(z)=-1+\frac{3}{2}\frac{(1+\omega)[c_{1}(1+z)]^{3(1+\omega)}}{1+[c_{1}(1+z)]^{3(1+\omega)}}\,, \label{5.22}
\end{equation}

\subsection{Model-II for $m=n=1$}

For $m=n=1$, the above field equations \eqref{5.5} \& \eqref{5.6} reduces to
\begin{equation}
	3(\mu-\lambda)H^{2}=\kappa^{2}\rho\,, \label{5.23}
\end{equation}
\begin{equation}
	(\mu-\lambda)(2\dot{H}+3H^{2})+\frac{\lambda k_{1}+\mu k_{2}}{3}H=-\kappa^{2}p\,, \label{5.24}
\end{equation}
For constant EoS parameter $\omega$, from the above equations \eqref{5.23} \& \eqref{5.24}, we obtain
\begin{equation}
	\dot{H}+\frac{3(1+\omega)}{2}H^{2}+\frac{\lambda k_{1}+\mu k_{2}}{6(\mu-\lambda)}H=0\,, \label{5.25}
\end{equation}
On integration of Eq.~\eqref{5.25}, we get
\begin{equation}
	H(t)=\frac{\frac{\lambda k_{1}+\mu k_{2}}{9(1+\omega)(\mu-\lambda)}}{c_{0}\exp\left(\frac{\lambda k_{1}+\mu _{2}}{6(\mu-\lambda)}t\right)-1}\,, \label{5.26}
\end{equation}
which gives the scale factor $a(t)$ as
\begin{equation}
	a(t)=c_{1}\left[c_{0}-\exp\left(-\frac{\lambda k_{1}+\mu _{2}}{6(\mu-\lambda)}t\right)\right]^{\frac{2}{3(1+\omega)}}\,, \label{5.27}
\end{equation}
where $c_{0}, c_{1}$ are arbitrary integrating constants.\\
Now, we can rewrite the field equations Eqs.~\eqref{5.23} \& \eqref{5.24} in the form of standard Friedmann field equations as
\begin{equation}
	3H^{2}=\kappa^{2}\rho+\kappa^{2}\rho_{(geom)}, \label{5.28}
\end{equation}
\begin{equation}
	2\dot{H}+3H^{2}=-\kappa^{2}p-\kappa^{2}p_{(geom)}, \label{5.29}
\end{equation}
where the effective dark energy density and pressure $\rho_{(geom)}$ and $p_{(geom)}$ coming from geometrical modifications are as given below:
\begin{equation}
	\rho_{(geom)}=\frac{3(1-\mu+\lambda)}{\kappa^{2}}H^{2}\,, \label{5.30}
\end{equation}
\begin{equation}
	p_{(geom)}=\frac{1}{\kappa^{2}}\left[\frac{\lambda k_{1}+\mu k_{2}}{3}H-(1-\mu+\lambda)(2\dot{H}+3H^{2})\right]\,, \label{5.31}
\end{equation}
The effective dark EoS parameter for model-II is obtained as
\begin{equation}
	\omega_{(geom)}=-1+\frac{(\lambda k_{1}+\mu k_{2})H-6(1-\mu+\lambda)\dot{H}}{9(1-\mu+\lambda)H^{2}}\,, \label{5.32}
\end{equation}
\begin{equation}
	H(z)=\frac{c_{0}(\lambda k_{1}+\mu k_{2})}{9(1+\omega)(\mu-\lambda)}[c_{1}(1+z)]^{\frac{3(1+\omega)}{2}}-\frac{\lambda k_{1}+\mu k_{2}}{9(1+\omega)(\mu-\lambda)}\,, \label{5.33}
\end{equation}
Equation \eqref{5.33} can be re-write as
\begin{equation}
	H(z)=\frac{H_{0}}{1-c_{0}c_{1}^{\frac{3(1+\omega)}{2}}}(1-c_{0}[c_{1}(1+z)]^{\frac{3(1+\omega)}{2}})\,, \label{5.34}
\end{equation}
where $H_{0}=\frac{(\lambda k_{1}+\mu k_{2})}{9(1+\omega)(\lambda-\mu)}[1-c_{0}c_{1}^{\frac{3(1+\omega)}{2}}]$ is the value of $H(z)$ at $z=0$.\\
Hence, the deceleration parameter $q(z)=-1+(1+z)\frac{H'}{H}$ with $H'=\frac{dH}{dz}$, is obtained as
\begin{equation}
	q(z)=-1+\frac{3}{2}\frac{c_{0}(1+\omega)[c_{1}(1+z)]^{\frac{3(1+\omega)}{2}}}{c_{0}[c_{1}(1+z)]^{\frac{3(1+\omega)}{2}}-1}\\, \label{5.35}
\end{equation}

\subsection{Model-III for $m=n=2$}

For $m=n=2$, the above field equations \eqref{5.5} \& \eqref{5.6} reduces to
\begin{equation}
	3(\mu-\lambda)H^{2}+\frac{\lambda k_{1}+\mu k_{2}}{2}H^{2}=\kappa^{2}\rho\,, \label{5.36}
\end{equation}
\begin{equation}
	(\mu-\lambda)(2\dot{H}+3H^{2})+\frac{\lambda k_{1}+\mu k_{2}}{2}\frac{2\dot{H}+7H^{2}}{3}=-\kappa^{2}p\,, \label{5.37}
\end{equation}
For constant EoS parameter $\omega$, the above equations \eqref{5.36} \& \eqref{5.37} gives
\begin{equation}
	\dot{H}+\frac{18(1+\omega)(\mu-\lambda)+(14+3\omega)(\lambda k_{1}+\mu k_{2})}{2[6(\mu-\lambda)+\lambda k_{1}+\mu k_{2}]}H^{2}=0\,, \label{5.38}
\end{equation}
On integration of Eq.~\eqref{5.38}, we get the Hubble parameter $H(t)$ as
\begin{equation}
	H(t)=\frac{1}{\frac{18(1+\omega)(\mu-\lambda)+(14+3\omega)(\lambda k_{1}+\mu k_{2})}{2[6(\mu-\lambda)+\lambda k_{1}+\mu k_{2}]}t+c_{0}}\,, \label{5.39}
\end{equation}
and the corresponding scale factor is obtained as
\begin{equation}
	a(t)=c_{1}\left[\frac{18(1+\omega)(\mu-\lambda)+(14+3\omega)(\lambda k_{1}+\mu k_{2})}{2[6(\mu-\lambda)+\lambda k_{1}+\mu k_{2}]}t+c_{0}\right]^{\frac{2[6(\mu-\lambda)+\lambda k_{1}+\mu k_{2}]}{18(1+\omega)(\mu-\lambda)+(14+3\omega)(\lambda k_{1}+\mu k_{2})}}\,, \label{5.40}
\end{equation}
where $c_{0}, c_{1}$ are arbitrary integrating constants.\\
Now, we can rewrite the field equations Eqs.~\eqref{5.36} \& \eqref{5.37} in the form of standard Friedmann field equations as
\begin{equation}
	3H^{2}=\kappa^{2}\rho+\kappa^{2}\rho_{(geom)}, \label{5.41}
\end{equation}
\begin{equation}
	2\dot{H}+3H^{2}=-\kappa^{2}p-\kappa^{2}p_{(geom)}, \label{5.42}
\end{equation}
where the effective dark energy density and pressure $\rho_{(geom)}$ and $p_{(geom)}$ coming from geometrical modifications are as given below:
\begin{equation}
	\rho_{(geom)}=\frac{6(1-\mu+\lambda)-(\lambda k_{1}+\mu k_{2})}{2\kappa^{2}}H^{2}\,, \label{5.43}
\end{equation}
\begin{equation}
	p_{(geom)}=\frac{1}{\kappa^{2}}\left[\frac{\lambda k_{1}+\mu k_{2}}{6}(2\dot{H}+7H^{2})-(1-\mu+\lambda)(2\dot{H}+3H^{2})\right]\,, \label{5.44}
\end{equation}
The effective dark EoS parameter for model-III is obtained as
\begin{equation}
	\omega_{(geom)}=-1+\frac{4(\lambda k_{1}+\mu k_{2})H^{2}+2[(\lambda k_{1}+\mu k_{2})-6(1-\mu+\lambda)]\dot{H}}{3[6(1-\mu+\lambda)-(\lambda k_{1}+\mu k_{2})]H^{2}}\,, \label{5.45}
\end{equation}
\begin{equation}
	H(z)=[c_{1}(1+z)]^{\frac{18(1+\omega)(\mu-\lambda)+(14+3\omega)(\lambda k_{1}+\mu k_{2})}{2[6(\mu-\lambda)+\lambda k_{1}+\mu k_{2}]}}\,, \label{5.46}
\end{equation}
Hence, the deceleration parameter $q(z)=-1+(1+z)\frac{H'}{H}$ with $H'=\frac{dH}{dz}$, is obtained as
\begin{equation}
	q(z)=-1+\frac{18(1+\omega)(\mu-\lambda)+(14+3\omega)(\lambda k_{1}+\mu k_{2})}{2[6(\mu-\lambda)+\lambda k_{1}+\mu k_{2}]}\,, \label{5.47}
\end{equation}

\section{Observational Constraints}

In this section, we utilize observational datasets to impose constraints on the model parameters of our derived model. To evaluate our derived model against observational datasets, we perform a Monte Carlo Markov Chain (MCMC) analysis using the emcee software, which is publically available at \cite{ref125}. The MCMC sampler analyzes the posterior distribution of the parameter space to constrain the model and cosmology parameters. It adjusts the parameter values within a range that is consistent with the prior distribution.\\

The Hubble parameter is a vital cosmological parameter for both theoretical and observational cosmologists who investigate the evolution of the cosmos. Initially, we employ Markov Chain Monte Carlo (MCMC) analysis to compare the Hubble function produced from the field equations with observed values of $H(z)$. This allows us to determine the best fit values of model parameters, along with their associated error ranges. This is facilitated by the accessibility of observed Hubble datasets $H(z)$ that include values of redshift $z$. In order to accomplish this, we use a total of 32 Hubble H(z) datasets that have been detected, each with its corresponding errors, as documented in references \cite{ref126,ref127,ref128,ref129,ref130,ref131,ref132,ref133}. We utilize the $\chi^{2}$-test formula in our investigation.
\begin{equation}\nonumber
	\chi^{2}(\phi)=\sum_{i=1}^{i=N}\frac{[(H_{ob})_{i}-(H_{th})_{i}]^{2}}{\sigma_{i}^{2}}
\end{equation}
where $N$ denotes the total number of data-points, $H_{ob},~H_{th}$, respectively, the observed and hypothesized datasets of $H(z)$ mentioned in Eqs.~\eqref{5.21}, \eqref{5.34} \& \eqref{5.47}, and standard deviations are displayed by $\sigma_{i}$. Here for the Model-I $\phi=(H_{0}, c_{1}, \omega)$ and for the Model-II $\phi=(H_{0}, c_{0}, c_{1}, \omega)$ and for Model-III $\phi=(H_{0}, \mu, \lambda, k_{1}, k_{2}, \omega)$.\\

For $\Lambda$CDM model, we consider the Hubble function as $H(z)=H_{0}\sqrt{\Omega_{m0}(1+z)^{3}+\Omega_{\Lambda0}}$ with $(\Omega_{m0}, \Omega_{\Lambda0})=(0.3, 0.7)$ at present.

\begin{table}[H]
	\centering
	\begin{tabular}{|c|c|c|c|}
		\hline
		Model         &	Parameter       & Prior            & Best Fit Value   \\
		\hline
		$\Lambda$CDM  &$H_{0}$          & $(50, 100)$      & $70.42_{-0.5945}^{+0.5986}$\\
		\hline
		              &$H_{0}$          & $(50, 100)$      & $67.799_{-0.10644}^{+0.10661}$\\
		Model-I       &$c_{1}$          & $(0, 1.5)$         & $0.75114_{-0.063149}^{+0.081214}$\\
		              &$\omega$         & $(-1, 1)$      & $0.012523_{-0.075138}^{+0.070474}$\\
		\hline
		              &$H_{0}$          &$(50,100)$       & $66.374_{-0.1043}^{+0.10421}$\\
		              &$c_{0}$          &$(-10,0)$           & $-1.6996_{-0.10319}^{+0.10508}$\\
		Model-II      &$c_{1}$          &$(0,1.5)$           & $0.7903_{-0.10466}^{+0.10241}$\\
		              &$\omega$         &$(-1,1)$        & $-0.0091266_{-0.077457}^{+0.094857}$\\
		\hline
				      &$H_{0}$          &$(50,100)$       & $65.7_{-0.099519}^{+0.097981}$\\
				      &$\mu$            &$(-2,3)$       & $0.39845_{-0.09666}^{+0.093595}$\\
				      &$\lambda$        &$(-2,30)$       & $1.7016_{-0.097638}^{+0.09977}$\\
		Model-III     &$k_{1}$          &$(-2,2)$           & $0.22972_{-0.091642}^{+0.088504}$\\
		              &$k_{2}$          &$(-2,3)$           & $0.71202_{-0.10131}^{+0.096535}$\\
		              &$\omega$         &$(-1,1)$        & $0.006333_{-0.089642}^{+0.088453}$\\
		 \hline
	\end{tabular}
	\caption{The MCMC Results in $H(z)$ datasets analysis.}\label{T1}
\end{table}

\begin{figure}[H]
	\centering
	\includegraphics[width=10cm,height=8cm,angle=0]{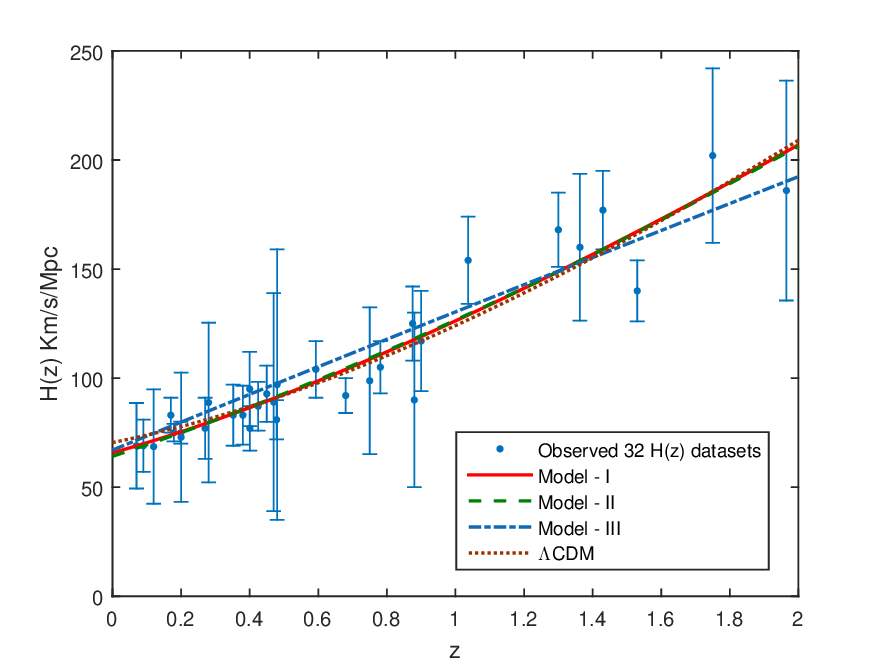}
	\caption{Hubble errorbar plots for Model-I, II, III and $\Lambda$CDM.}
\end{figure}

\begin{figure}[H]
	\centering
	\includegraphics[width=6cm,height=8cm,angle=0]{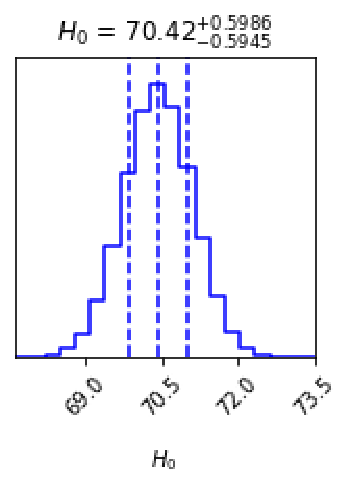}
	\caption{The contour plot of $H_{0}$ at $1-\sigma, 2-\sigma$ and $3-\sigma$ confidence level in MCMC analysis of $H(z)$ datasets for $\Lambda$CDM model.}
\end{figure}

\begin{figure}[H]
	\centering
	\includegraphics[width=10cm,height=10cm,angle=0]{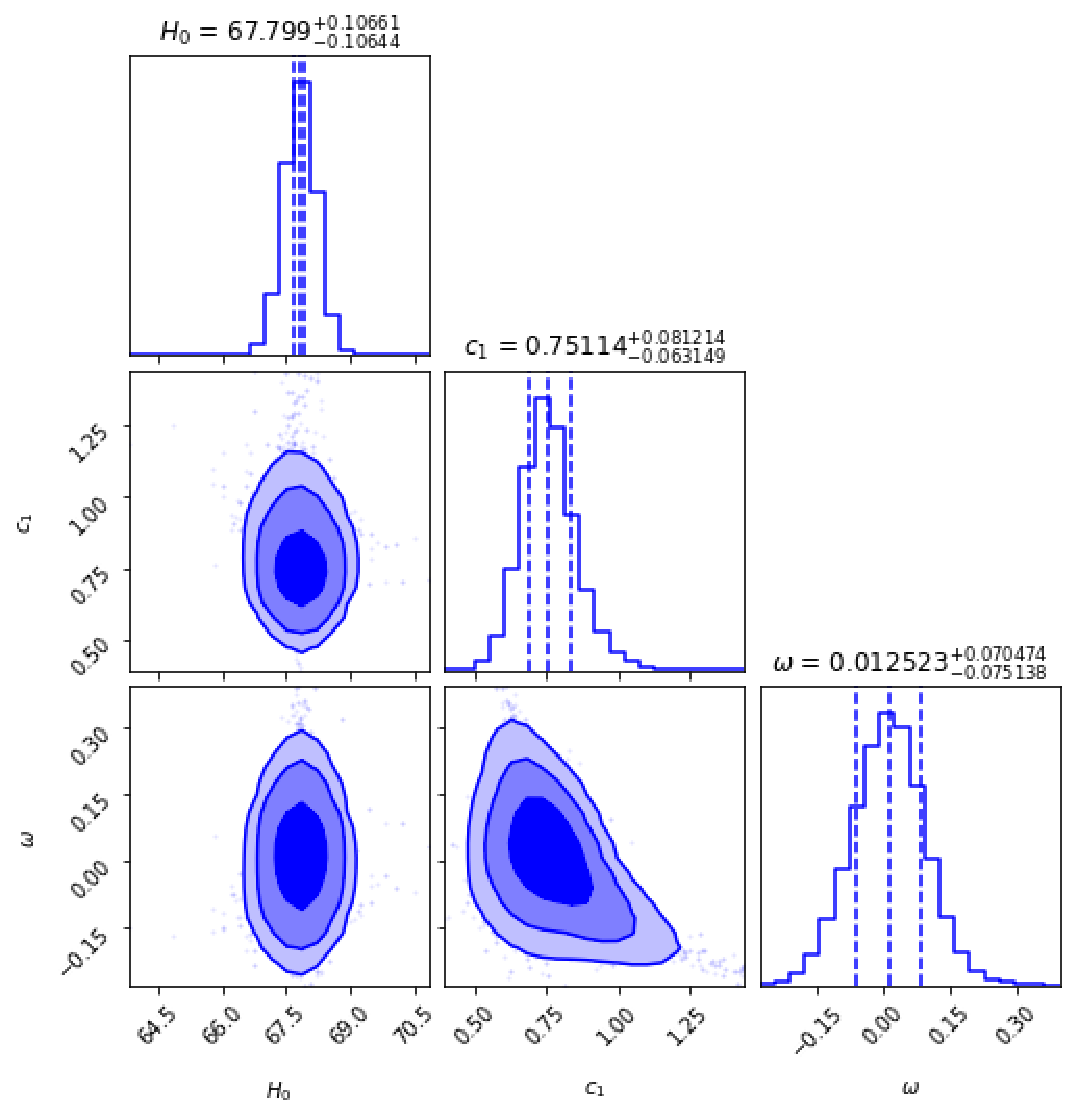}
	\caption{The contour plots of $H_{0}, c_{1}, \omega$ at $1-\sigma, 2-\sigma$ and $3-\sigma$ confidence level in MCMC analysis of $H(z)$ datasets for Model-I.}
\end{figure}

\begin{figure}[H]
	\centering
	\includegraphics[width=10cm,height=10cm,angle=0]{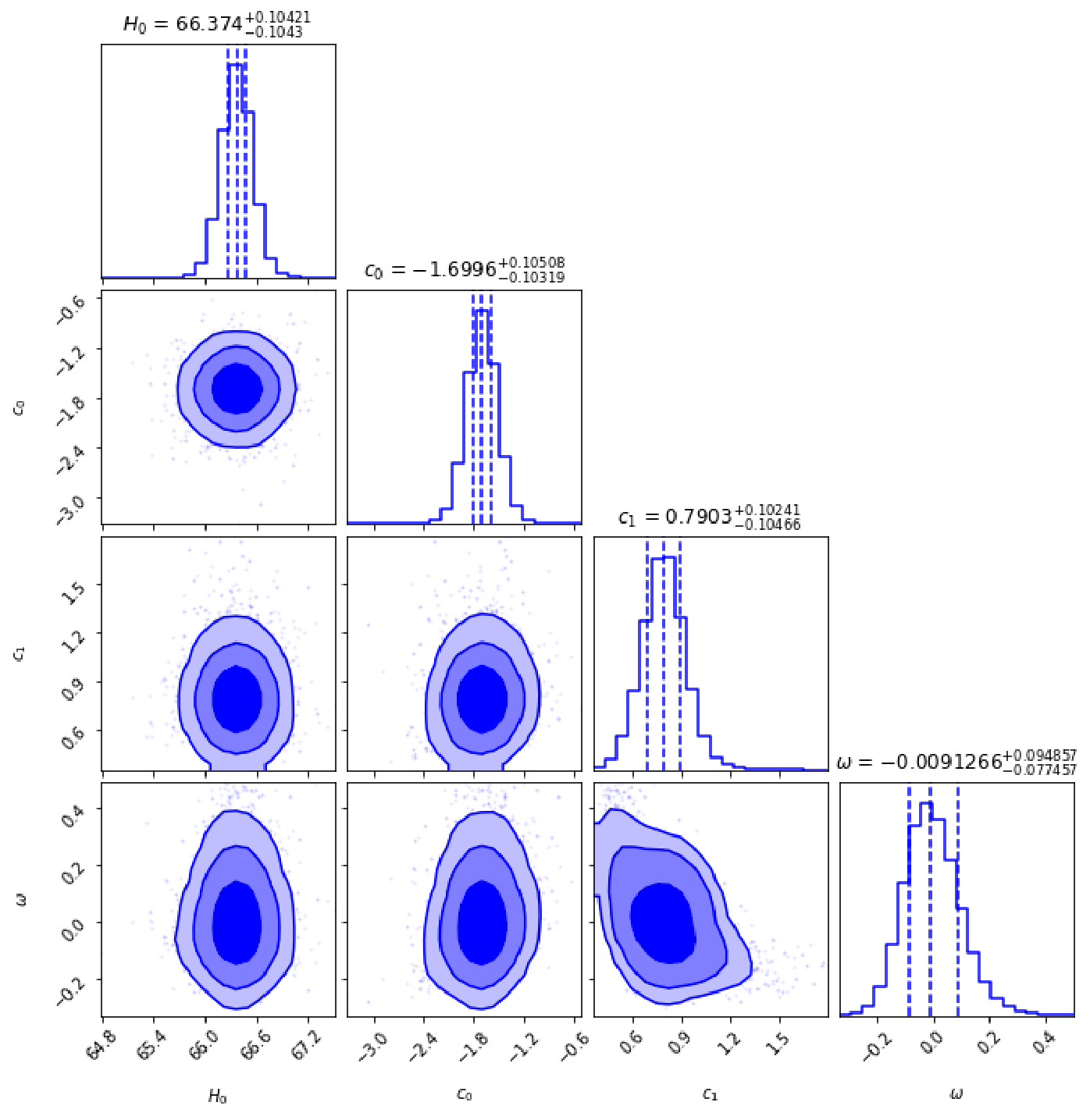}
	\caption{The contour plots of $H_{0}, c_{0}, c_{1}, \omega$ at $1-\sigma, 2-\sigma$ and $3-\sigma$ confidence level in MCMC analysis of $H(z)$ datasets for Model-II.}
\end{figure}

\begin{figure}[H]
	\centering
	\includegraphics[width=10cm,height=10cm,angle=0]{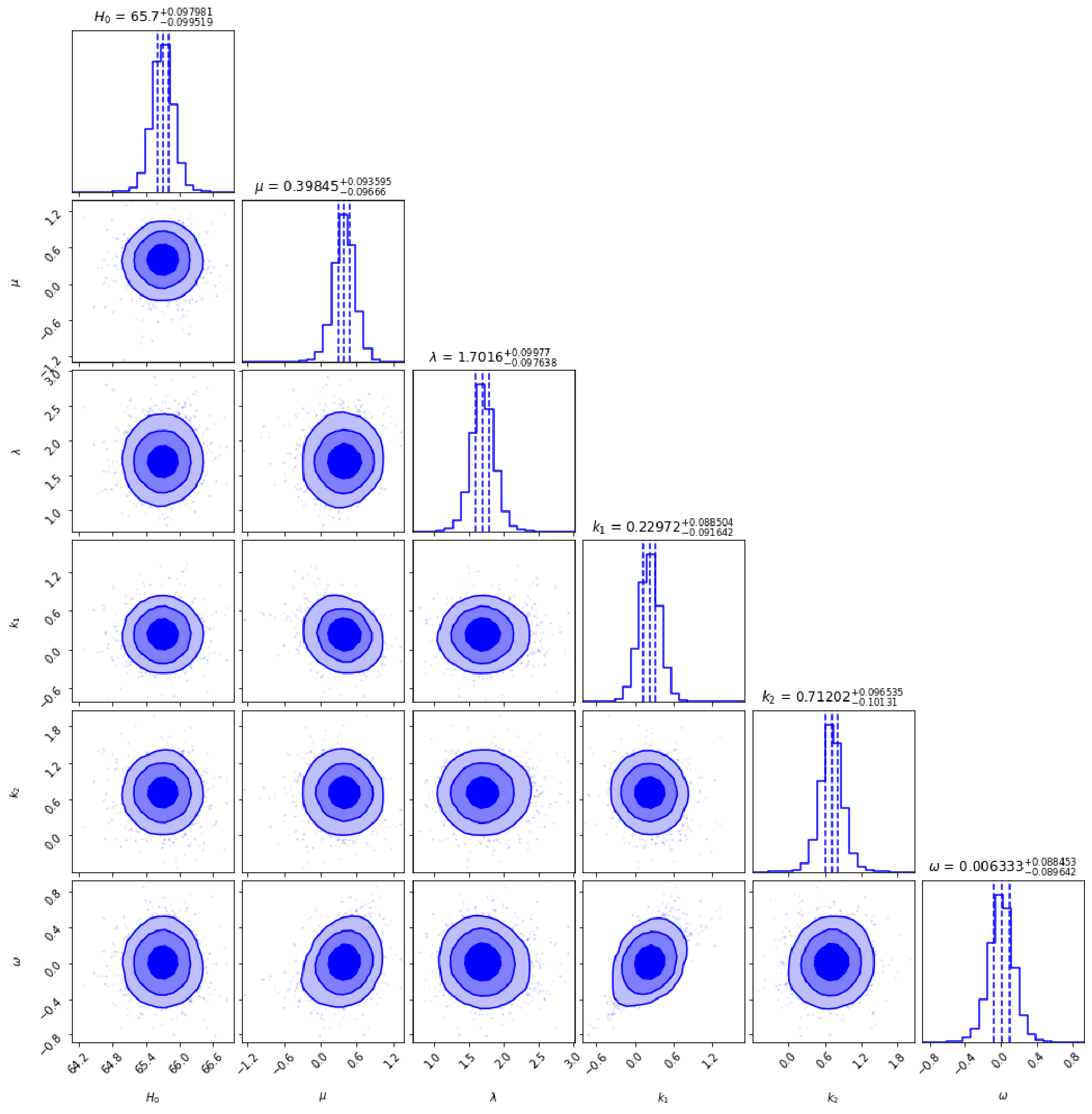}
	\caption{The contour plots of $H_{0}, \mu, \lambda, k_{1}, k_{2}, \omega$ at $1-\sigma, 2-\sigma$ and $3-\sigma$ confidence level in MCMC analysis of $H(z)$ datasets for Model-III.}
\end{figure}
The mathematical expressions for the Hubble functions obtained in three derived Models Model-I, II, III are shown in above Eqs.~\eqref{5.21}, \eqref{5.34} \& \eqref{5.47}, respectively, and the best fit shapes are shown in figure 1 for the best fit values as mentioned in Table 1. Figure 2 depicts the likelihood plot obtained from MCMC analysis of $\Lambda$CDM model with $H(z)$ datasets. Figure 3 depicts the contour plots of $H_{0}, c_{1}, \omega$ at $1-\sigma, 2-\sigma$ and $3-\sigma$ confidence level for $H(z)$ datasets for model-I. Figure 4 shows the contour plots of $H_{0}, c_{0}, c_{1}, \omega$ at $1-\sigma, 2-\sigma$ and $3-\sigma$ confidence level in MCMC analysis of $H(z)$ datasets for Model-II. Figure 5 represents the contour plots of $H_{0}, \mu, \lambda, k_{1}, k_{2}, \omega$ at $1-\sigma, 2-\sigma$ and $3-\sigma$ confidence level in MCMC analysis of $H(z)$ datasets for Model-III. In all the three models, we have estimated the Hubble constant values as $H_{0}=67.799_{-0.10644}^{+0.10661}, 66.374_{-0.1043}^{+0.10421}, 65.7_{-0.099519}^{+0.097981}$ Km/s/Mpc, respectively, while for $\Lambda$CDM model it is estimated as $H_{0}=70.42_{-0.5945}^{+0.5986}$ Km/s/Mpc which are consistent with the recent observed values of $H_{0}$.

\section{Result discussions}

The mathematical expression of deceleration parameter $q(z)$ for model-I is given in Eq.~\eqref{5.22}, and its geometrical interpretation is shown in figure 6. We observe that $q(z)$ is an increasing function of $z$ and it depicts a transition from decelerating to accelerating phase. For the best fit values of model parameters, we have estimated the present value of deceleration parameter $q_{0}=-0.5513$ for the model-I that reveals the present accelerating phase of the expanding universe. To investigate the phase transition line, the transition redshift $z_{t}$ is calculated as
\begin{equation}\label{7.1}
	z_{t}=\frac{1}{c_{1}}\left[ \frac{2}{1+3\omega}\right]^{\frac{1}{3(1+\omega)}}-1 
\end{equation}
The estimated value of the transition redshift is obtained as $z_{t}=0.6524$ and also, one can see that this transition value depends upon the parameters $c_{1}, \omega$. Its dependency on $c_{1}, \omega$ reveals the importance of this model. The present value of the deceleration parameter $q_{0}<0$ indicates that our derived model-I is accelerating phase at present. In the figure 6, the negative values of the redshift represent the future universe while the past universe is depicted by the positive values of $z$. The present phase of the universe is obtained at $z=0$. Thus, from the figure 6, one can see that model-I is decelerating in early time and it is in accelerating phase at present and future time as $q\to-1$ as $z\to-1$. The line $q=0$ represents the transition line of the evolution of the universe. Also, the present value of Hubble parameter is estimated as $H_{0}=67.799_{-0.10644}^{+0.10661}$ Km/s/Mpc.\\

For model-II, the mathematical expression for deceleration parameter is given in the Eq.~\eqref{5.35}, and its geometrical behaviour is shown in figure 6. We observe that the $q(z)$ function is an increasing function of $z$ and it evolves with a signature-flipping point (transition redshift) that can be derived from Eq.~\eqref{5.35} as
\begin{equation}\label{7.2}
	z_{t}=\frac{1}{c_{1}}\left[\frac{-2}{c_{0}(1+3\omega)}\right]^{\frac{2}{3(1+\omega)}}-1 
\end{equation}
The transition redshift is estimated as $z_{t}=0.4258$ which are consistent with recent estimated values. The present value of deceleration parameter is estimated as $q_{0}=-0.1872$ that depicts that model-II is in accelerating phase at present. From figure 6, one can observe that $q\to-1$ as $z\to-1$ that deals the accelerating phase of the universe at late-time universe while $q\to0.5$ as $z\to\infty$ that shows the past decelerating phase of the universe. The estimated present value of Hubble parameter for model-II is obtained as $H_{0}=66.374_{-0.1043}^{+0.10421}$ Km/s/Mpc. For the model-III, we obtain a constant deceleration parameter and estimated value of $q=-0.0098$ for the best fit model parameters values that depicts the accelerating model of the universe. The best fit value of Hubble parameter for the model-III is obtained as $H_{0}=65.7_{-0.099519}^{+0.097981}$ Km/s/Mpc.\\

We have also, investigated the deceleration behaviour of $\Lambda$CDM model by taking the Hubble function as $H(z)=H_{0}\sqrt{\Omega_{m0}(1+z)^{3}+\Omega_{\Lambda0}}$. We have estimated the present value of Hubble constant as $H_{0}=70.42_{-0.5945}^{+0.5986}$ Km/s/Mpc with present deceleration parameter $q_{0}=-0.55$. The transition redshift for $\Lambda$CDM model is obtained as $z_{t}=0.6711$. From figure 6, one can observe that both model-I, II are similar in behaviour with $\Lambda$CDM model but model-I is more closed to $\Lambda$CDM in comparison of model-II.
\begin{figure}[H]
	\centering
	\includegraphics[width=10cm,height=8cm,angle=0]{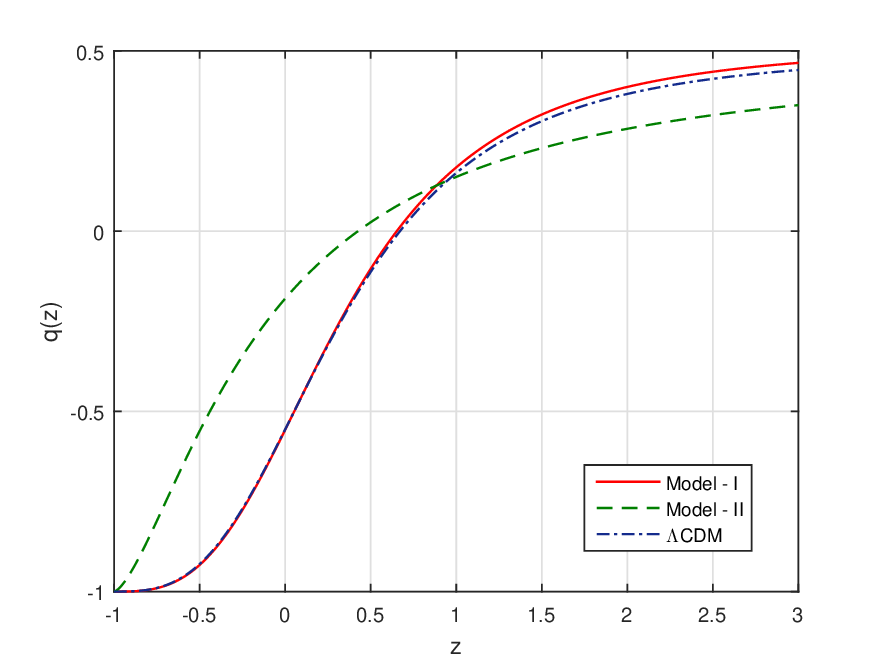}
	\caption{The plots of deceleration parameter $q(z)$ for the Model-I, II and $\Lambda$CDM versus $z$ for the best fit values of model parameters.}
\end{figure}
Now, we have tried to find the best fit values of model parameters $\lambda, \mu, k_{1}$ and $k_{2}$ for the model-I, II by using the relation $H_{0}=\sqrt{\frac{6(\mu-\lambda)}{\lambda k_{1}+\mu k_{2}}}\sqrt{1+c_{1}^{3(1+\omega)}}$ for model-I and $H_{0}=\frac{(\lambda k_{1}+\mu k_{2})}{9(1+\omega)(\lambda-\mu)}[1-c_{0}c_{1}^{\frac{3(1+\omega)}{2}}]$ for the model-II at $z=0$. Using these relations, we have plotted the $\frac{\lambda}{\mu}$ for model-I and $\frac{\mu}{\lambda}$ for model-II as function of $k_{1}, k_{2}$. These plots are shown in figure 7a \& 7b respectively. From figure 7a \& 7b, we have selected the value of $\lambda$ \& $\mu$ such that for which we get $\rho_{(geom)}\ge0$ and hence, we have estimated as $\lambda=-0.003098, \mu=0.1, k_{1}=100, k_{2}=3.1$ for model-I and $\lambda=0.1, \mu=0.07315, k_{1}=20, k_{2}=46$ for model-II. We have used these values of model parameters $\lambda, \mu, k_{1}, k_{2}$ for rest of the analysis of cosmological properties of the model-I, II.
\begin{figure}[H]
	\centering
	a.\includegraphics[width=7cm,height=6cm,angle=0]{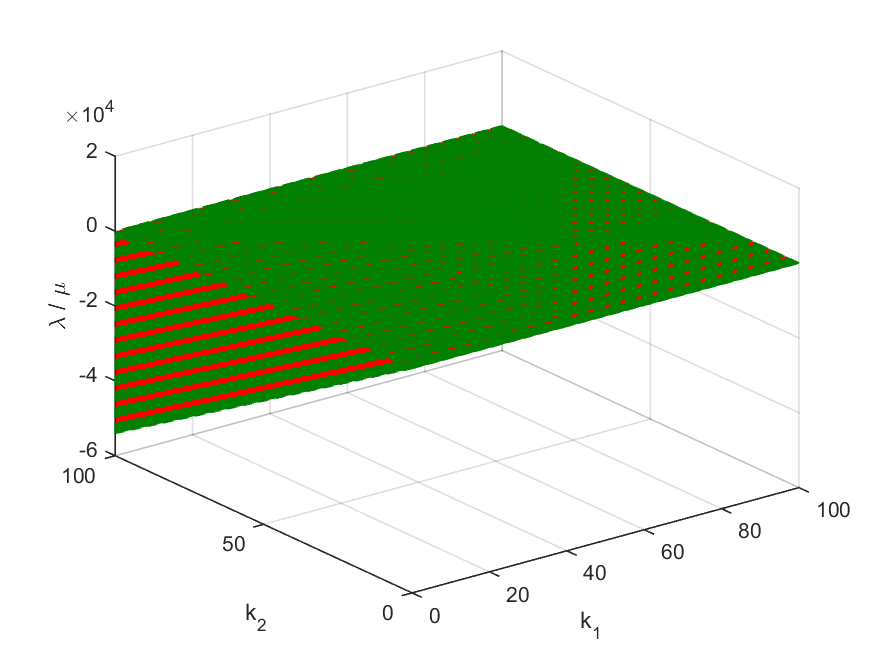}
	b.\includegraphics[width=7cm,height=6cm,angle=0]{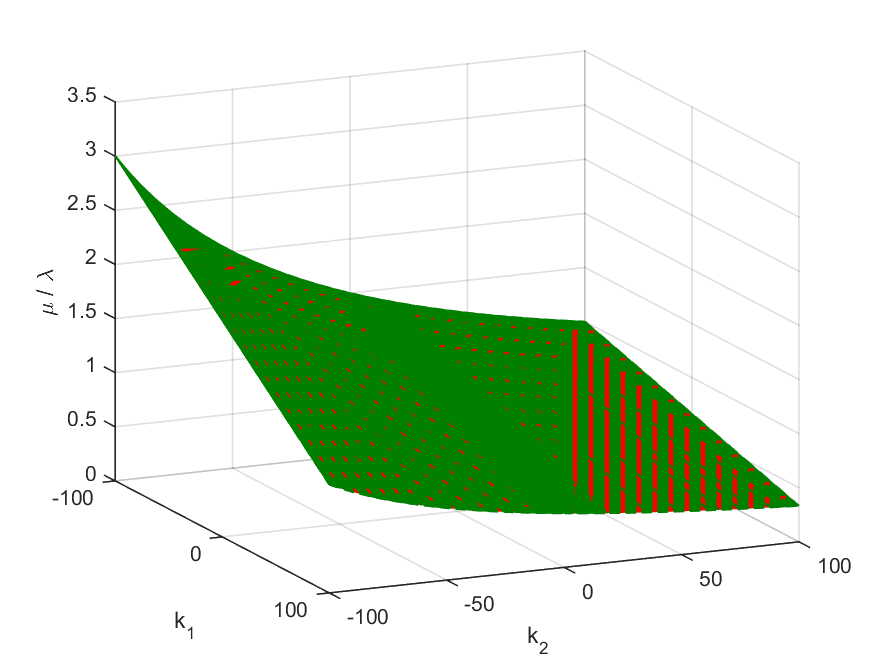}
	\caption{The plots of $\lambda/\mu$ and $\mu/\lambda$ versus $k_{1}, k_{2}$, respectively for model-I nad model-II.}
\end{figure}

\begin{figure}[H]
	\centering
	a.\includegraphics[width=7cm,height=6cm,angle=0]{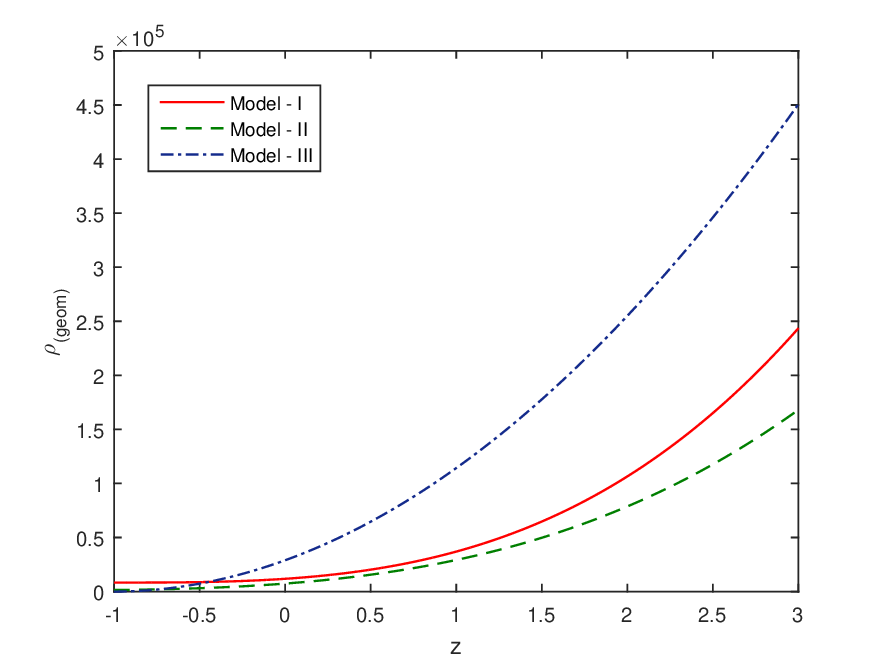}
	b.\includegraphics[width=7cm,height=6cm,angle=0]{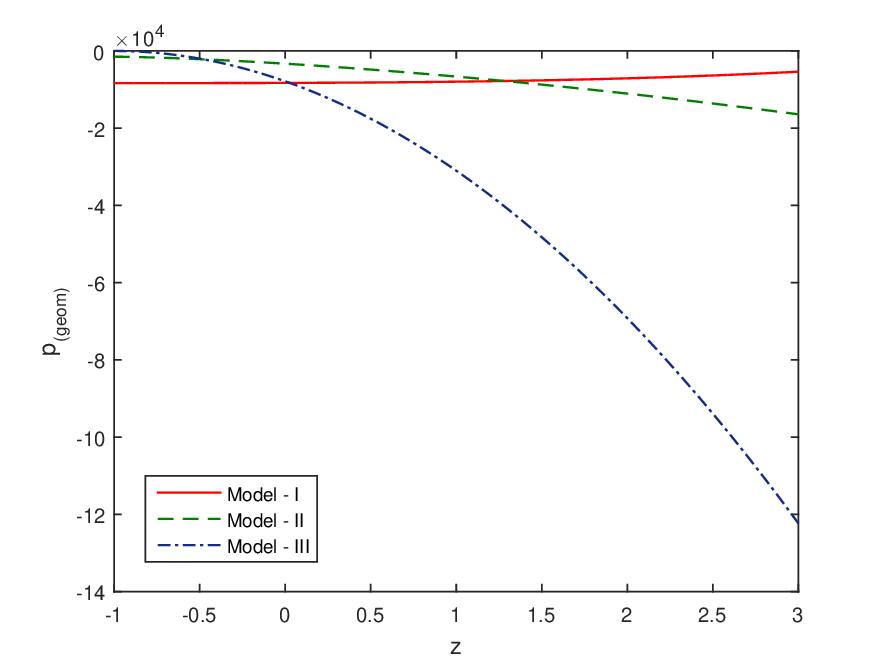}
	\caption{The plots of dark energy density $\rho_{(geom)}$ and dark pressure $p_{(geom)}$ versus $z$, respectively.}
\end{figure}
Figure 8a \& 8b represent the behaviour of dark energy density $\rho_{(geom)}$ and dark pressure $p_{(geom)}$, respectively for model-I, II, III and the mathematical expressions of $\rho_{(geom)}$ and $p_{(geom)}$ are shown in Eqs.~\eqref{5.15}, \eqref{5.30}, \eqref{5.43} and \eqref{5.16}, \eqref{5.31}, \eqref{5.44}, respectively. From figure 8a \& 8b, one can see that $\rho_{(geom)}\ge0$ and $p_{(geom)}\le0$ for all $z$. It confirms that geometrical modification creates dark energy term which has a high negative pressure for all three models, that may cause the acceleration in the expansion of the universe.
\begin{figure}[H]
	\centering
	\includegraphics[width=10cm,height=8cm,angle=0]{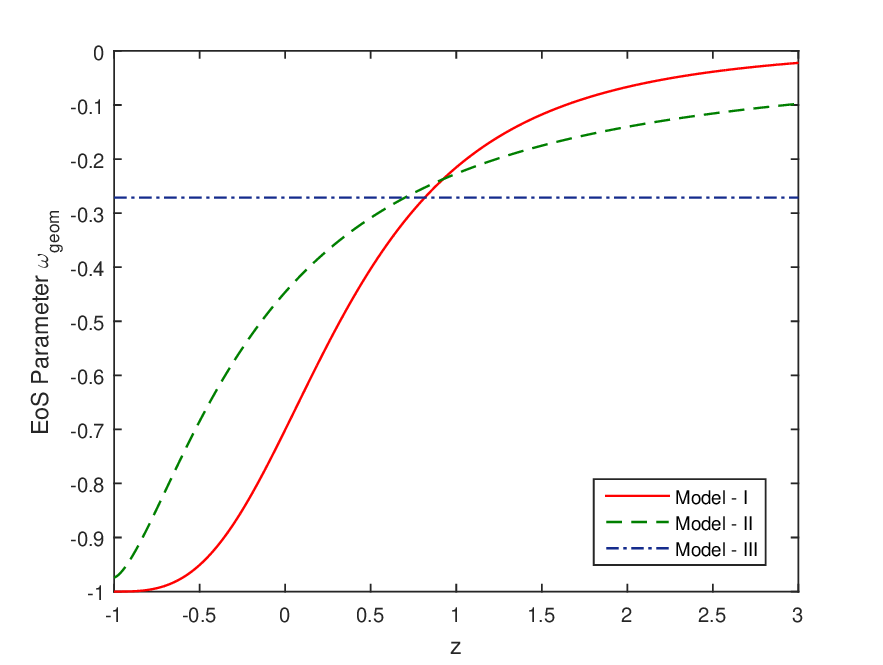}
	\caption{The plots of effective EoS parameter $\omega_{(geom)}$ versus $z$.}
\end{figure}

Figure 9, represents the geometrical evolution of effective EoS parameter $\omega_{(geom)}$ over $z$ for all three models and the mathematical expressions are given in Eqs.~\eqref{5.17}, \eqref{5.32} \& \eqref{5.45}, respectively for model-I, II, III. From figure 9, one can observe that $\omega_{(geom)}$ is an increasing function of $z$ for model-I, II while it is constant for model-III. We have estimated the present values of effective EoS parameter $\omega_{(geom)}=-0.7009, -0.4463, -0.2715$, respectively for all three models with best fit model parameters values. From figure 9, one can see that the model-I and model-II behaves just like quintessence dark energy models and late-time it tends to $\Lambda$CDM models as $\omega_{(geom)}\to-1$ as $z\to-1$.

\section{Conclusions}

This research examines precise cosmological models within the framework of Myrzakulov $F(T,Q)$ gravity, also known as Myrzakulov gravity-III (MG-III) as described in [arXiv:1205.5266], while taking into account empirical constraints. The MG-III gravity is a form of unification between two established theories of gravity, specifically, the $F(T)$ gravity and the $F(Q)$ gravity. The field equations of the MG-III theory are derived by treating the metric tensor and the general affine connection as separate and unrelated variables. Next, we examine the specific scenario where the $F(T,Q)$ function that defines the metric-affine models discussed before is linear, meaning $F(T,Q)=\lambda T+\mu Q$. We examine the linear scenario and analyze a Friedmann-Lema\^{i}tre-Robertson-Walker backdrop in order to explore cosmological characteristics and potential uses. We have derived three precise solutions of the modified field equations for various scenarios involving $T$ and $Q$. These solutions are obtained in terms of the Hubble function $H(t)$ and the scale factor $a(t)$. To validate these solutions, we have applied observational constraints using the Hubble $H(z)$ datasets and conducted an MCMC analysis. We have conducted an investigation on the deceleration parameter $q(z)$, as well as the effective equation of state (EoS) parameters. Furthermore, we have performed a comparison analysis of all three models in relation to the $\Lambda$CDM model. The main features of the derived models are as follows:
\begin{itemize}
	\item All the derived models are exact solutions of the modified Einstein's field equations. In Model-I \& II, we have found hyperbolic scale factor while model-III governs by a power-law scale solution.
	
	\item Model-I and Model-II are found as transit phase expanding universe model which are decelerating in past and accelerating at present and in late-time.
	
	\item We have found the best fit values of Hubble constant as $H_{0}=67.799_{-0.10644}^{+0.10661}, 66.374_{-0.1043}^{+0.10421}, 65.7_{-0.099519}^{+0.097981}$ Km/s/Mpc for all three models, respectively, which are consistent with recent observed values of Hubble constant.
	
	\item We have found the present values of deceleration parameter as $q_{0}=-0.5513, -0.1872, -0.0098$, respectively for all three models, that shows all the models are currently in accelerating expansion phase. The value $q_{0}$ corresponding to $\Lambda$CDM model is obtained as $q_{0}=-0.55$ with Hubble constant $H_{0}=70.42_{-0.5945}^{+0.5986}$ Km/s/Mpc.
	
	\item The transition redshift for the transit phase models I, II are found as $z_{t}=0.6524, 0.4258$ while for $\Lambda$CDM model it is found as $z_{t}=0.6711$. All these results are compatible with region $0.3<z_{t}<1$.
	
	\item We have estimated the values of model parameters $\lambda$ and $\mu$ as $\lambda=-0.003098, \mu=0.1, k_{1}=100, k_{2}=3.1$ for model-I and $\lambda=0.1, \mu=0.07315, k_{1}=20, k_{2}=46$ for model-II.
	
	\item We have found the present value of effective EoS parameter as $\omega_{(geom)}=-0.7009, -0.4463, -0.2715$, respectively for all three models with best fit model parameters values. Model-I and model-II behaves just like quintessence dark energy model and late-time it tends to $\Lambda$CDM scenarios.
\end{itemize}

Thus, the models in non-Riemannian geometry may explain the late-time accelerating phase of the expanding universe without introducing cosmological constant $\Lambda$. This type of theory of gravity is more generalization of $f(T)$ and $f(Q)$ gravity theory, and this type of models need more investigation and hence, it may be interesting for readers in this field.

\section*{Acknowledgments}

The work was supported by the Ministry of Education and Science of the Republic of Kazakhstan, Grant
AP14870191.
\section{Data Availability Statement}
No data associated in the manuscript.
\section{Statements and Declarations}
\subsection*{Funding and/or Conflicts of interests/Competing interests}
The author of this article has no conflict of interests. The author have no competing interests to declare that are relevant to the content of this article. Authors have mentioned clearly all received support from the organization for the submitted work.
\newpage
	
\end{document}